\documentclass[
    ,final            
  ]
  {aipproc}

\layoutstyle{6x9}

\newcommand{\be}{\begin{equation}}
\newcommand{\ee}{\end{equation}}
\newcommand{\bea}{\begin{eqnarray}}
\newcommand{\eea}{\end{eqnarray}}
\newcommand{\bean}{\begin{eqnarray*}}
\newcommand{\eean}{\end{eqnarray*}}

\begin{document}

\title{Understanding $I=2$  $\pi\pi$  Interaction}

\author{B.S.Zou}{
  address={Institute of High Energy Physics, CAS, Beijing 100039, China}
}

\author{F.Q.Wu}{
  address={Institute of High Energy Physics, CAS, Beijing 100039, China}
}

\author{L.Li}{
  address={Institute of High Energy Physics, CAS, Beijing 100039, China}
  ,altaddress={Peking University, Beijing 100087, China}
}

\author{D.V.Bugg}{
  address={Queen Mary College, London, UK}
}

\begin{abstract}
A correct understanding and description of the $I=2$ $\pi\pi$
S-wave interaction is important for the extraction of the $I=0$
$\pi\pi$ S-wave interaction from experimental data and for
understanding the $I=0$ $\pi\pi$ S-wave interaction theoretically.
With t-channel $\rho$, $f_2(1270)$ exchange and the
$\pi\pi\to\rho\rho\to\pi\pi$ box
  diagram contribution, we reproduce the $\pi \pi$ isotensor S-wave and D-wave
  scattering phase shifts and inelasticities up to 2.2 GeV
  quite well in a K-matrix formalism.
\end{abstract}

\maketitle


\section{Introduction}

Much attention has been paid to the isospin I=0 $\pi\pi$ S-wave
interaction due to its direct relation to the $\sigma$ particle
and the scalar glueball candidates. However, to really understand
the isoscalar $\pi\pi$ S-wave interaction, one must first
understand the isospin I=2 $\pi\pi$ S-wave interaction due to the
following two reasons:  (1) There are no known s-channel
resonances and less coupled channels in I=2 $\pi\pi$ system, so it
is much simpler than the I=0 $\pi\pi$ S-wave interaction; (2) To
extract I=0 $\pi\pi$ S-wave phase shifts from experimental data on
$\pi^+\pi^-\to\pi^+\pi^-$ and $\pi^+\pi^-\to\pi^0\pi^0$ obtained
by $\pi N\to\pi\pi N$ reactions, one needs an input of the I=2
$\pi\pi$ S-wave interaction. While the $I=2$ $\pi\pi$ S-wave
interaction can be extracted from the pure $I=2$
$\pi^\pm\pi^\pm\to\pi^\pm\pi^\pm$ reactions, the $I=0$ $\pi\pi$
S-wave interaction can only be extracted from
$\pi^+\pi^-\to\pi^+\pi^-$ and $\pi^+\pi^-\to\pi^0\pi^0$ reactions
which are mixture of $I=0$ and $I=2$ contributions. The relation
between the $\pi^+\pi^-\to\pi^+\pi^-$, $\pi^0\pi^0$ S-wave
amplitudes and the isospin decoupled amplitudes is as the
following:
\begin{eqnarray}
 T_{s}(+-,+-) &=& T_s^{I=0}/3+T_s^{I=2}/6 ,\label{tpm}\\
 T_{s}(+-,00) &=& T_s^{I=0}/3-T_s^{I=2}/3. \label{t00}
\end{eqnarray}
The $T_s^{I=0}$ was usually extracted from $T_{s}(+-,+-)$ and
$T_{s}(+-,00)$ information by assuming some kind of $T_s^{I=2}$
amplitude.

Up to now, experimental information on the I=2 $\pi\pi$ scattering
mainly came from $\pi^+p\to\pi^+\pi^+n$ \cite{Hoogland77} and
$\pi^-d\to\pi^-\pi^-pp$ \cite{Durusoy73,Cohen} reactions. The main
features for the I=2 $\pi\pi$ S-wave phase shifts $\delta_0^2$ and
inelasticities $\eta_0^2$ are: (1) the $\delta_0^2$ goes down more
and more negative as the $\pi\pi$ invariant mass increases from
$\pi\pi$ threshold up to 1.1 GeV; (2) the $\delta_0^2$ starts to
increase for energies above about 1.1 GeV; (3) the $\eta_0^2$
starts to deviate from 1 for energies above 1.1 GeV. The first
feature can be well explained by the t-channel $\rho$ exchange
force \cite{Isgur,lilong,Zou94,Speth} while the effect of
t-channel scalar exchange is extremely small and can be neglected
\cite{Speth}. Due to the relative poor quality of the I=2 $\pi\pi$
scattering data above 1.1 GeV, the other two features are usually
overlooked. In this work \cite{Wu}, we show in a K-matrix
formalism \cite{lilong,lilong2} that these two features can be
well reproduced by the t-channel $f_2(1270)$ exchange and the
$\pi\pi$-$\rho\rho$ coupled-channel effect, respectively.

Recently, the $\pi^+ \pi^-\to \pi^+ \pi^- $ scattering from the
old $\pi N$ scattering experiments with both unpolarized\cite{CM}
and polarized targets\cite{Becker} has been re-analyzed\cite{BSZ,
Kaminski} in combination with new information from $p\bar p$ and
other experiments. The $\pi^+\pi^-\to\pi^0\pi^0$ scattering has
also been studied by E852\cite{E852}, GAMS\cite{GAMS}
Collaborations and analyzed \cite{Achasov}. In Refs.\cite{CM,BSZ},
a scattering length formula for I=2 $\pi \pi$ S-wave was used; in
Ref.\cite{Achasov} another empirical parametrization was used.
These parametrizations give similar phase shifts up to 1.1 GeV,
but differ at higher energies. All these previous analyses have
ignored the inelastic effects in the $I=2$ channel. We will
demonstrate that the correct description of the $I=2$ S-wave
interaction has significant impact on the extraction of the $I=0$
$\pi\pi$ S-wave amplitude for energies above 1.1 GeV.

\section{Theoretical framework}

We follow the $K$-matrix formalism as in
Refs.\cite{lilong,lilong2}. For the two-channel case, the
two-dimensional $K$ matrix and phase space $\rho(s)$ matrix are
\be K=\left (
\begin{array}{ccc}
K_{11}  &  K_{12}  \\
K_{12}  &  K_{22}
\end{array} \right ),\hspace{1cm}
\rho(s)=\left (
\begin{array}{ccc}
\rho_1(s)  &  0  \\
0  &  \rho_2(s)
\end{array} \right ),
\ee with i=1,2 representing $\pi\pi$ and $\rho\rho$ channel,
respectively.   \be
T_{11}=\frac{K_{11}-i\rho_2(K_{11}K_{22}-K_{12}K_{21})}{1-i\rho_1
K_{11}-i\rho_2 K_{22} -\rho_1\rho_2(K_{11}K_{22}-K_{12}K_{21})},
\ee Ignoring the interaction between $\rho\rho$, we have
$K_{22}=0$; then
\be T_{11}=\frac{K_{11}+i K_{12} \rho_2 K_{21}}{1-i
\rho_1(K_{11}+iK_{12} \rho_2 K_{21} )}, \ee
where $iK_{12} \rho_2 K_{21}$ corresponds to the contribution of
the $\pi\pi\to\rho\rho\to\pi\pi$ box diagram.

In order to obtain $K_{11}$,  we incorporate the $t$-channel
$f_2(1270)$ contribution into the $t$-channel $\rho$ exchange term
by the Dalitz-Tuan method \cite{lilong,BSZ}. \be
K_{11}=\frac{K_{\rho}(s)+K_{f_2}(s)}{1-\rho_1^2(s)K_{\rho}(s)
K_{f_2}(s)}.\ee

 The amplitudes for the t-channel $\rho$ and $f_2(1270)$ meson exchange
without considering the vertex form factor were given in
Ref.\cite{Zou94} for studying $I=0$ $\pi\pi$ scattering. In this
work, we include a form factor of conventional monopole type to
take into account the off-shell behavior of the exchanged mesons:
$ F(q^2)=(\Lambda^2-m^2)/(\Lambda^2-q^2)$ with (m, q) the mass and
four-vector momentum, respectively, of exchanged mesons, and
$\Lambda$ the cutoff parameter to be determined by experimental
data. The resulted $K_\rho$ and $K_{f_2}$ are given in our paper
\cite{Wu}.

\begin{figure}[htbp]
\includegraphics[width=10cm,height=2cm]{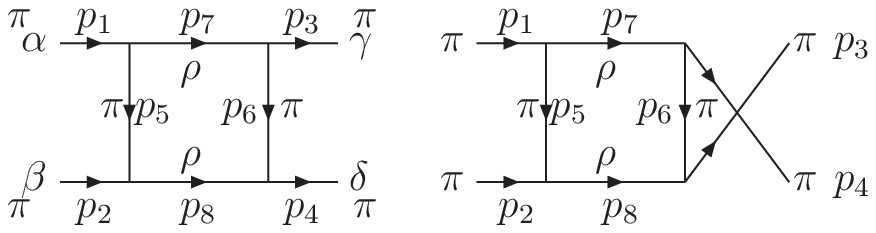}
 \caption{The $\pi\pi\to\rho\rho\to\pi\pi$ box diagrams}
 \label{block1}
\end{figure}

For energies above the $\rho\rho$ threshold, the inelastic effect
should be taken into account in the $I=2$ $\pi\pi$ channel. Note
that unlike $I=0$ $\pi\pi$ channel, the $I=2$ $\pi\pi$ channel
does not couple to the $K\bar K$ and $\omega\omega$ channels due
to isospin conservation. The $\pi\pi$ channel couples to the
$\rho\rho$ channel by the t-channel $\pi$ exchange as shown in
Fig.\ref{block1}.

 Assuming the on-shell approximation \cite{Lu} for the $\rho\rho$
intermediate state, the amplitude for the box diagrams are
calculated and projected to the S- and D-wave \cite{Wu}. It is
related to the K-matrix by $ T_{box}^{I=2} = i K_{12} \rho_2
K_{12}$. For the phase space factor $\rho_2$, the width of the
$\rho$ meson is taken into account in our calculation.

\begin{figure}
\begin{minipage}[t]{0.45\textwidth}
\includegraphics[width=7cm, height=7cm]{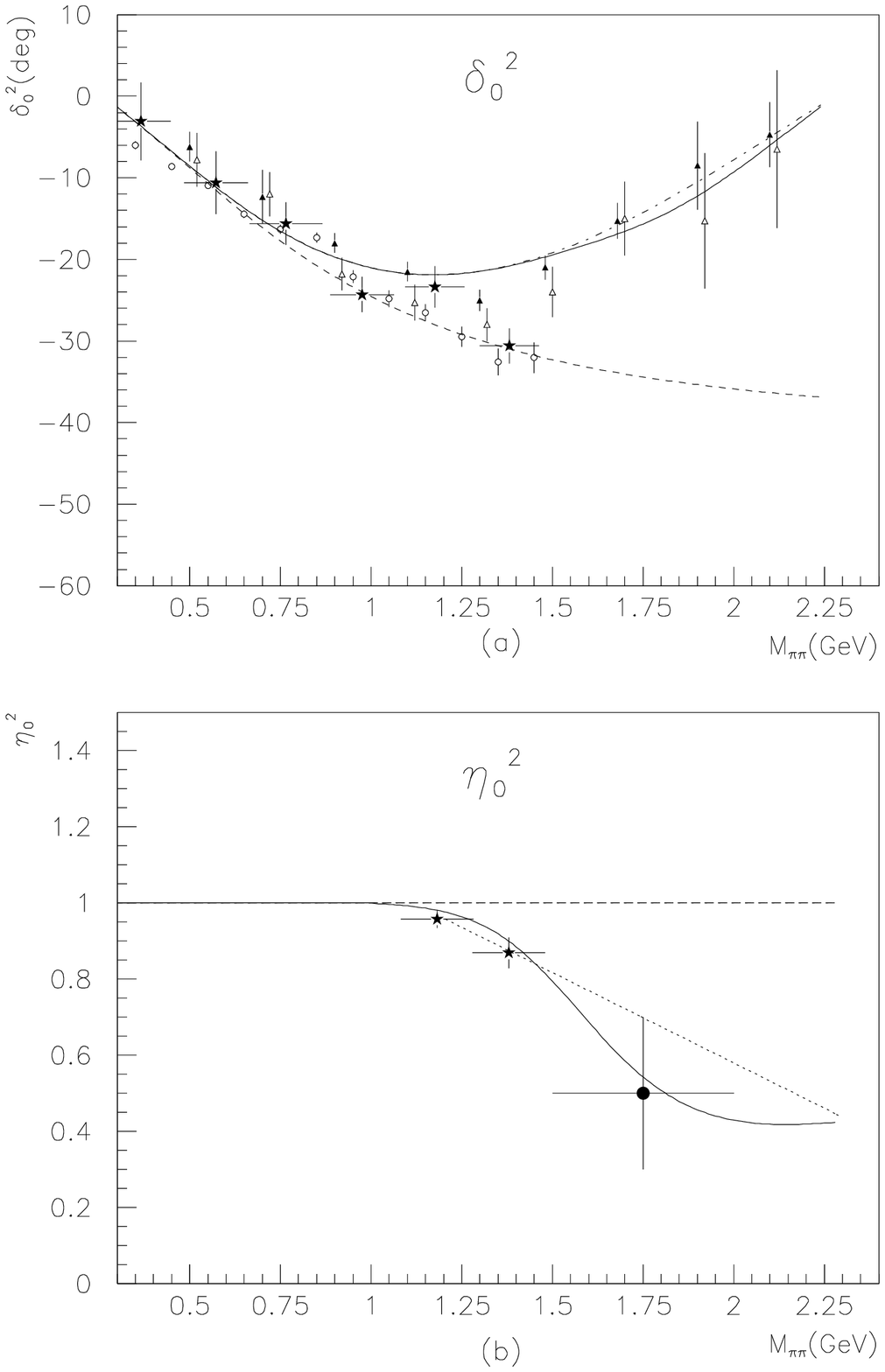}
\end{minipage}
\hfill
  \begin{minipage}[t]{0.5\textwidth}
\includegraphics[width=7cm,height=7cm]{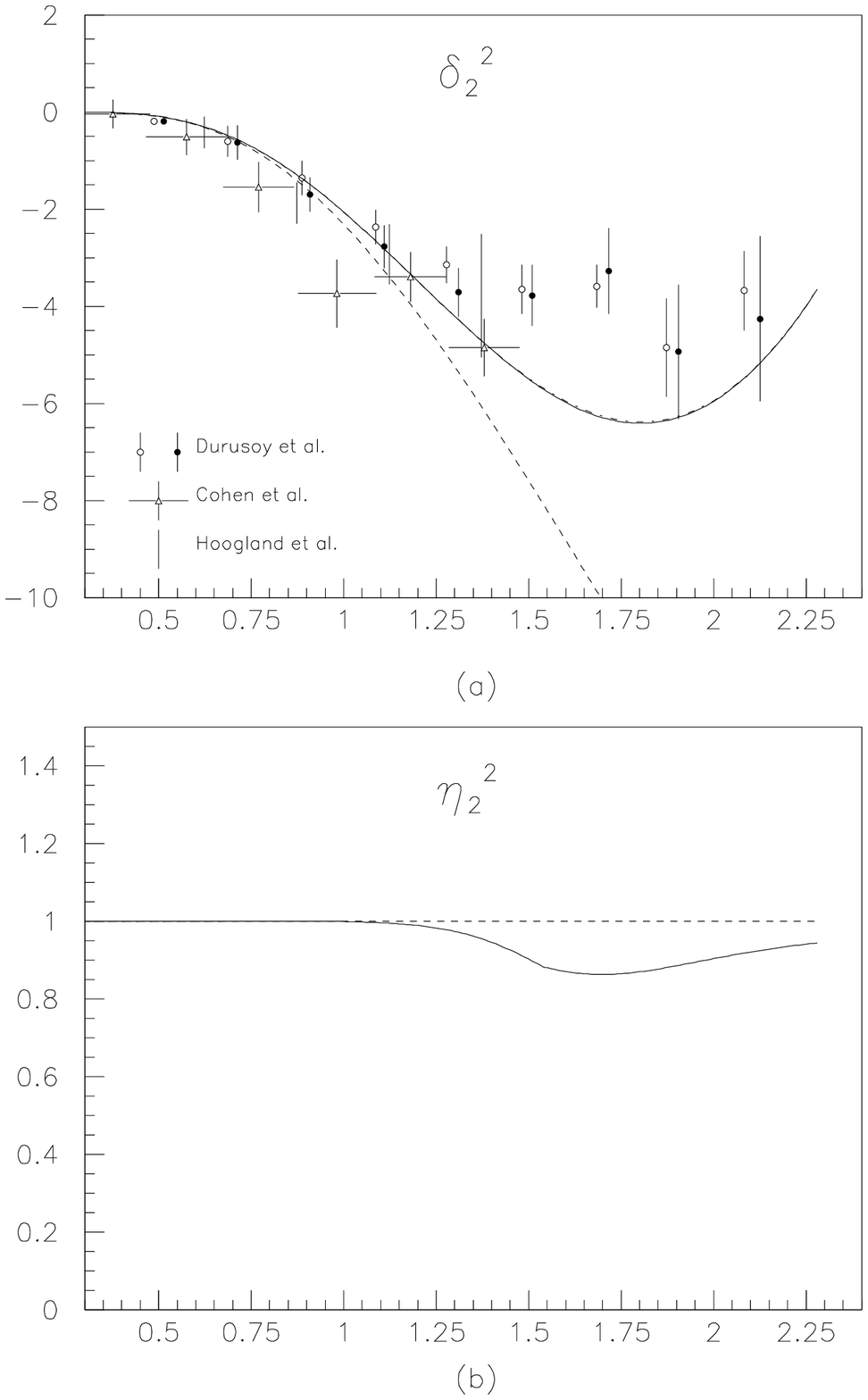}
\caption{The $I=2$ $\pi\pi$ $S$-wave ($\delta_0^2$, $\eta_0^2$)
and $D$-wave ($\delta_2^2$, $\eta_2^2$) phase shifts and
inelasticities. Data are from
Ref.\cite{Hoogland77,Durusoy73,Cohen}. The solid curves represent
the total contribution of $\rho$, $f_2$ exchange and  the box
diagram; dot-dashed curves from $\rho $ and $f_2$ exchange; dashed
curves from t-channel $\rho$ exchange only.} \label{phase}
\end{minipage}
\end{figure}

\section{Numerical results and discussion}

\ \ \ \ From the formalism given above, we get I=2 $\pi\pi$ S-wave
and D-wave phase shifts and inelasticities as shown in
Fig.\ref{phase}.  The t-channel $\rho$ exchange alone (dashed
lines) reproduces the phase shifts for energies up to 1.1 GeV very
well with the form factor parameter $\Lambda_{\rho\pi\pi}=1.5$ GeV
\cite{lilong}, but underestimates the phase shifts at higher
energies. The inclusion of the t-channel $f_2(1270)$ exchange
(dot-dashed lines) increases the phase shifts especially for
energies above 1 GeV and can reproduce the phase shift data very
well with the form factor parameter $\Lambda_{f_2\pi\pi}=1.7$ GeV.
However they only contribute to the elastic scattering and cannot
produce the inelasticities for energies above 1 GeV.

The experimental information on the inelasticities is scarce for
the $I=2$ $\pi\pi$ scattering. Two data points were given by
Ref.\cite{Cohen} for energies $1\sim 1.5$ GeV. For energies
$1.5\sim 2$ GeV, Ref.\cite{Durusoy73} estimated to be $0.5\pm 0.2$
for the $\eta^2_0$ in one solution and assumed
$\eta_0^2=1.53-0.475m_{\pi\pi}$ (GeV/$c^2$) for another solution.
The two solutions gave similar results for the $I=2$ $\pi\pi$
S-wave phase shifts. For the $I=2$ $\pi\pi$ D-wave scattering, the
inelasticity could not be measured well and was assumed to be
$\eta^2_2=1$ for the extraction of the $\delta^2_2$.

\begin{figure}[htbp]
\begin{minipage}[t]{0.5\textwidth}
\includegraphics[width=7cm,height=4cm]{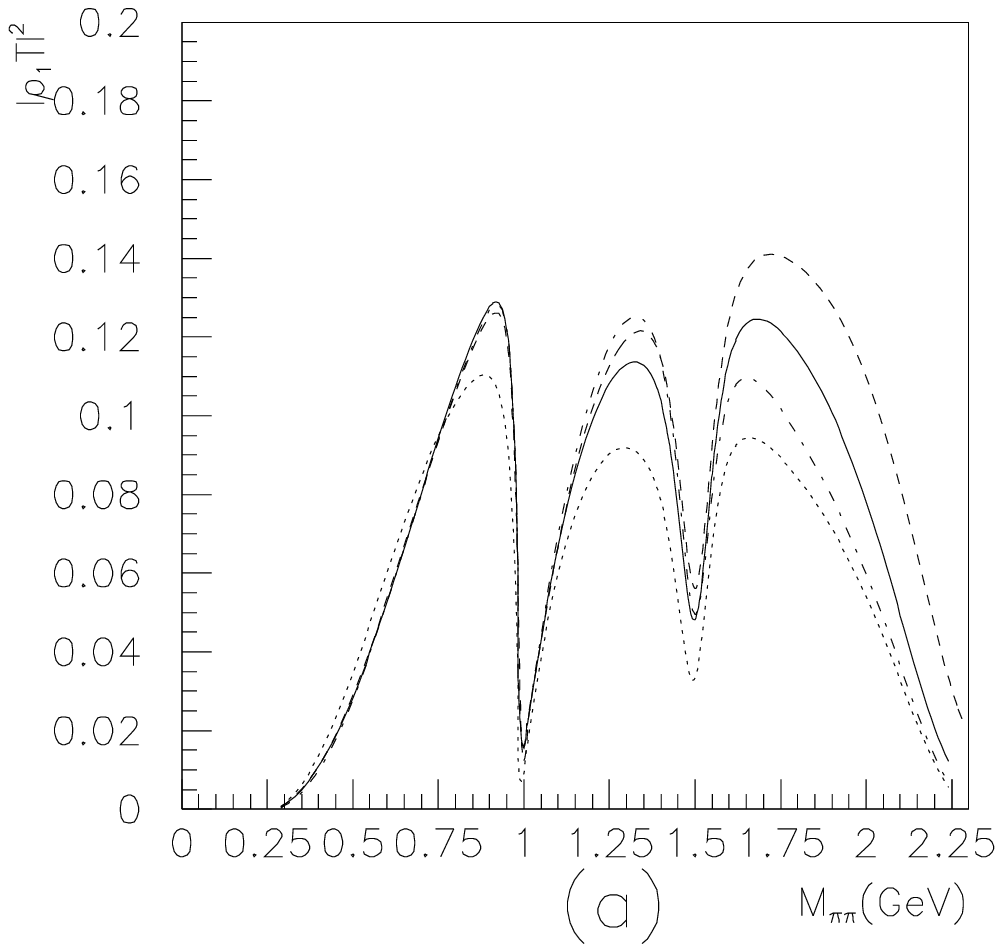}
\end{minipage}
\hfill
  \begin{minipage}[t]{0.5\textwidth}
\includegraphics[width=7cm,height=4cm]{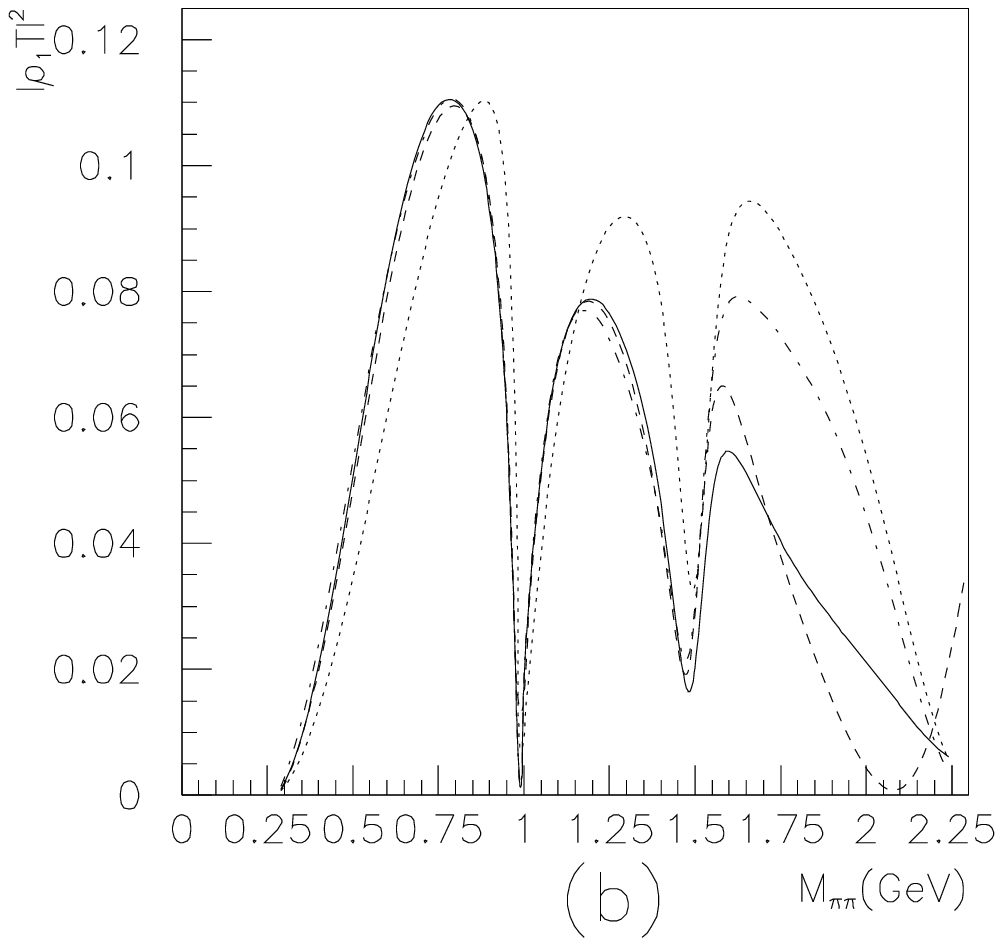}
\end{minipage}
\caption{Full amplitudes squared of $\pi^+\pi^-\to\pi^+\pi^- $ (a)
and $\pi^+\pi^-\to\pi^0\pi^0$ (b). The lines correspond to using
$T^{I=0}_s$ from Ref.\cite{lilong} plus various input of
$T^{I=2}_s$: $T^{I=2}_s=0$ (dotted lines); $T^{I=2}_s$ from
Refs.\cite{lilong,CM} (dot-dashed lines); $T^{I=2}_s$ from
Ref.\cite{Achasov} (dashed lines); $T^{I=2}_s$ from this work
(solid line). }
 \label{sampl}
\end{figure}

Although there are only three data points with large error bars
for the inelasticity parameter $\eta^2_0$ of the $I=2$ $\pi\pi$
S-wave scattering, it is clear that the inelastic effect may be
significant around 1.6 GeV. In order to reproduce this
inelasticity, it is necessary to consider the
$\pi\pi\leftrightarrow\rho\rho$ coupling channel effect. We find
that including contribution from the $\pi\pi\to\rho\rho\to\pi\pi$
box diagram in our K-matrix formalism, the $I=2$ S-wave
inelasticity data can be very well reproduced without introducing
any more free parameter as shown in Fig.\ref{phase}. The same
diagram also predicts a broad shallow dip around 1.7 GeV for the
inelasticity of the $I=2$ D-wave scattering. Assuming $\eta^2_2=1$
for energies around 1.7 GeV may bias the $\delta^2_2$ data around
this energy. This may be the reason that the $\delta^2_2$ data
around 1.7 GeV has the largest discrepancy with our theoretical
result. The box diagram has little influence to the phase shifts
although it produces large inelasticity.

Inspired by the new claim of a pentaquark state \cite{Nakano}, we
also explored the possibility of including an $I=2$ s-channel
resonance to reproduce the $I=2$ $\pi\pi$ S-wave scattering data
instead of using the t-channel $f_2$ exchange and the
$\pi\pi\to\rho\rho\to\pi\pi$ box diagram, but failed to reproduce
the data. While the $\eta^2_0$ needs the resonance with mass
around 1.6GeV, the $\delta^2_0$ needs the mass above 2.3 GeV with
a much broader width.

To demonstrate the significance of possible impact of the $I=2$
input for the extraction of $I=0$ $\pi\pi$ amplitude, we calculate
the full S-wave amplitudes for  $\pi^+ \pi^- \to \pi^+ \pi^- $ and
$\pi^+ \pi^- \to \pi^0 \pi^0 $ according to
Eqs.(\ref{tpm},\ref{t00}) with $T^{I=0}_s$ from Ref.\cite{lilong}
plus various $T^{I=2}_s$ inputs. The corresponding full S-wave
amplitudes squared ($|\rho_1T|^2$) are shown in Fig.\ref{sampl}.
The dotted lines are the results with $T^{I=2}_s=0$. The
dot-dashed lines correspond to the scattering length formula used
in Refs.\cite{lilong,CM,BSZ}, which is similar to the result by
considering only the t-channel $\rho$ exchange contribution. The
dashed lines use the new empirical $T^{I=2}_{s}$ formula of
Ref.\cite{Achasov}, which is similar to the result by considering
t-channel $\rho$ and $f_2$ exchange contributions, but ignoring
the inelasticity caused by the $\pi\pi\leftrightarrow\rho\rho$ box
diagram contribution. The solid lines are results with our
$T^{I=2}_{s}$ including the t-channel $\rho, f_2$ exchange and the
box diagram. It is clear that $T^{I=2}_s$ has significant
contribution to the amplitudes for $\pi^+\pi^-\to\pi^+\pi^-$ and
$\pi^+\pi^-\to\pi^0\pi^0$ processes, hence has significant impact
on the extraction of the $T^{I=0}_s$ amplitude. Previous inputs of
$T^{I=2}_s$ give similar results as our new $T^{I=2}_s$ for
energies below 1.1 GeV, but differ from ours significantly for
higher energies. The inclusion of both t-channel $f_2$ exchange
and $\pi\pi\to\rho\rho\to\pi\pi$ box diagram contribution is
important.

In summary, the basic features of $I=2$ $\pi\pi$ scattering phase
shifts and inelasticities can be well reproduced by the t-channel
meson ($\rho$,$f_2$) exchange and the
$\pi\pi\leftrightarrow\rho\rho$ coupled-channel effect in the
K-matrix formalism. The t-channel $\rho$
  exchange provides repulsive negative phase shifts while the
  t-channel $f_2(1270)$ gives an attractive force to increase the phase shifts
  for $\pi \pi$ scattering above 1 GeV, and the coupled-channel box diagram
  causes the inelasticities. A correct description of the $I=2$ $\pi\pi$
scattering has significant impact on the extraction of the $I=0$
scattering amplitudes from $\pi^+\pi^-\to\pi^+\pi^-$ and
$\pi^+\pi^-\to\pi^0\pi^0$ data, especially for energies above 1.2
GeV . A re-analysis of these data with our new description of the
$I=2$ $\pi\pi$ scattering will be carried out as our next step.


\begin{theacknowledgments}
  The work is partly supported by CAS
Knowledge Innovation Project (KJCX2-SW-N02) and the National
Natural Science Foundation of China.
\end{theacknowledgments}


\bibliographystyle{aipproc}   


\IfFileExists{\jobname.bbl}{}
 {\typeout{}
  \typeout{******************************************}
  \typeout{** Please run "bibtex \jobname" to optain}
  \typeout{** the bibliography and then re-run LaTeX}
  \typeout{** twice to fix the references!}
  \typeout{******************************************}
  \typeout{}
 }

\end{document}